\documentclass[conference, 9pt]{IEEEtran}
\IEEEoverridecommandlockouts
\usepackage{cite}
\usepackage{amsmath,amssymb,amsfonts}
\usepackage[noend]{algorithmic}
\usepackage{graphicx}
\usepackage{url} 
\usepackage{textcomp}
\usepackage{xcolor}
\usepackage{balance}
\usepackage{booktabs}
\usepackage{algorithm}
\usepackage{booktabs,tabularx}
\usepackage{pifont}     

\makeatletter
\newcommand\fs@ruledtight{\fs@ruled\def\@fs@post{\kern2pt\hrule\relax\kern-16pt}}
\makeatother
\floatstyle{ruledtight}
\restylefloat{algorithm}
\usepackage[a4paper, total={184mm,239mm}]{geometry}
\def\BibTeX{{\rm B\kern-.05em{\sc i\kern-.025em b}\kern-.08em
    T\kern-.1667em\lower.7ex\hbox{E}\kern-.125emX}}

\AtBeginDocument{%
  \setlength{\abovedisplayskip}{1.0ex plus 2.8pt minus 1.4pt}%
  \setlength{\belowdisplayskip}{1.0ex plus 2.8pt minus 1.4pt}%
  \setlength{\abovedisplayshortskip}{0pt plus 2.8pt}%
  \setlength{\belowdisplayshortskip}{1.0ex plus 2.8pt minus 1.4pt}%
}

\begin{document}

\title{ReLay: Bidirectional Layout--Code Translation and Library Evolution for Reusable and Process-Portable Analog Design%
\thanks{Preprint. Under review.
This work has been submitted to the IEEE for possible publication.
Copyright may be transferred without notice, after which this version
may no longer be accessible.}
\looseness=-1
}

\author{
\IEEEauthorblockN{
Minxing Chu\IEEEauthorrefmark{1},
Quan Cheng\IEEEauthorrefmark{1},
Anhang Li\IEEEauthorrefmark{2},
Martins Atilla\IEEEauthorrefmark{1},\\
Junyi Luo\IEEEauthorrefmark{1},
Xinting Jiang\IEEEauthorrefmark{1},
Ruichen Qi\IEEEauthorrefmark{1},
Mehdi Saligane\IEEEauthorrefmark{1}
}
\IEEEauthorblockA{
\IEEEauthorrefmark{1}Brown University, Providence, RI, USA\\
\IEEEauthorrefmark{2}University of Michigan, Ann Arbor, MI, USA\\
\texttt{\{minxing\_chu,martins\_atilla,junyi\_luo,xinting\_jiang,ruichen\_qi,mehdi\_saligane\}@brown.edu}\\
\texttt{ivin.chengquan@gmail.com, anhangli@umich.edu}
}
}

\maketitle

\begin{abstract}
Analog layout generators provide a programmable interface for design reuse and process migration, but their coverage is limited by handcrafted libraries. The designer knowledge embedded in complete layouts remains implicit in their geometry, making it difficult to capture in reusable generator libraries. We present ReLay, a bidirectional layout–code framework that recovers generator programs from existing layouts, enabling cross-PDK portability and parametric resizing of existing designs. Constructions that recur across recovered programs are promoted to library functions, so the library evolves as layouts accumulate. We also introduce the first layout--code benchmark: 1,050 hand-drawn layouts from 27 circuit families across three open-source PDKs, ranging from primitive cells to full chips. Learned functions account for 53–66\% of layout-building calls in held-out layouts, demonstrating reuse beyond the learning corpus. Frontier LLM agents use ReLay to produce new layouts at 64–77\% lower API cost per completed task than direct geometry generation, with comparable completion rates and electrical quality.
\looseness=-1
\end{abstract}

\section{Introduction}

Analog layout remains one of the least automated steps of integrated circuit design, despite being a long-standing goal of electronic design automation. A digital design passes from its register-transfer-level description
through synthesis, placement, and routing to a finished layout with little human
intervention. Analog layout, in contrast, must consider circuit-specific matching, symmetry,
parasitics, and process rules, so
design quality depends on designer judgment and tape-out experience. Constraint-driven systems such as ALIGN~\cite{Kunal2019ALIGN} and
MAGICAL~\cite{Xu2019MAGICAL} synthesize a layout from a netlist through
placement and routing guided by extracted constraints, while generator
frameworks such as BAG~\cite{Chang2018BAG2}, LAYGO~\cite{Han2021LAYGO}, and
gLayout~\cite{Hammoud2024GLayoutLLM,Hammoud2024GLayoutRL} let designers write
parameterized programs that generate layouts from design parameters.

Generator frameworks are appealing because they express layout as a program that is modular and hierarchical and can be versioned, reviewed, and reused with standard software tooling. The rise of large language models
(LLMs) strengthens this appeal, as LLMs can now solve complex programming
tasks~\cite{Chen2021Codex} and have begun to design non-trivial analog
circuits~\cite{Lai2025AnalogCoder} and
layouts~\cite{Liu2025LayoutCopilot,Hammoud2024GLayoutLLM}. Because LLMs work far
more fluently with code than with geometry~\cite{Chu2026CodeBehindLayout}, a generator language offers them a natural interface for creating and editing layouts. Among generator
frameworks, gLayout~\cite{Hammoud2024GLayoutLLM}, on which we instantiate this work, brings this approach to
open-source PDKs. It describes analog layouts as structured, modular, and
hierarchical Python code, in which parametric primitives such as transistors, via
stacks, and guard rings are placed and wired through named ports. Its defining
feature is a PDK-agnostic design, in which every primitive is written once
against an abstract description of the process, so that the same program
generates the layout in each supported PDK. These primitives and composition rules make coding a layout much like writing a software program. 
\looseness=-1

\begin{figure}[t]
  \centering
  \includegraphics[width=\columnwidth]{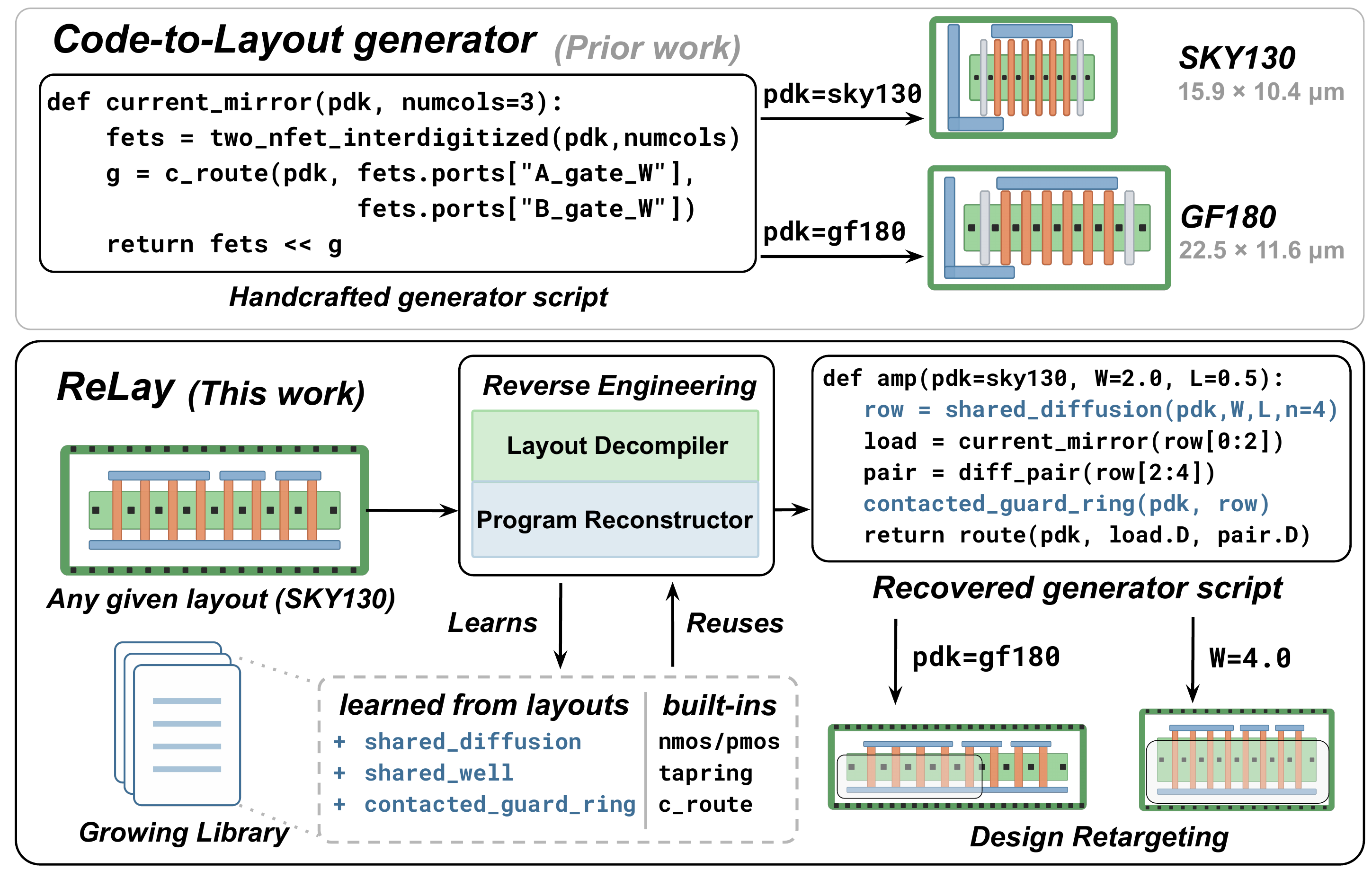}
  \vspace{-18pt}
  \caption{Unidirectional and bidirectional generator workflows.}
  \label{fig:intro-compare}
  \vspace{-13pt}
\end{figure}

However, a handcrafted generator framework can express only what its library
provides. Its primitives and composite cells, for example, only cover typical structures, so an unusual interdigitation pattern or a custom guard ring
has to be drawn polygon by polygon. The library grows only when its maintainers
add functions one at a time for the structures they foresee, and its coverage
falls short of the variety found in real designs. In addition, writing a
generator demands both layout expertise, from device matching to design rules,
and the programming skill to express that expertise as parameterized code, which
makes the framework hard to adopt for designers used to drawing layouts by hand.
As a result, the design knowledge held in many finished layouts can
hardly reach the library, and generator code remains scarce. This scarcity also limits LLMs, whose skill at writing generators depends heavily on how many example programs they can learn from and how much of the design space the library covers.

\begin{figure*}[t]
  \centering
  \includegraphics[width=0.95\textwidth]{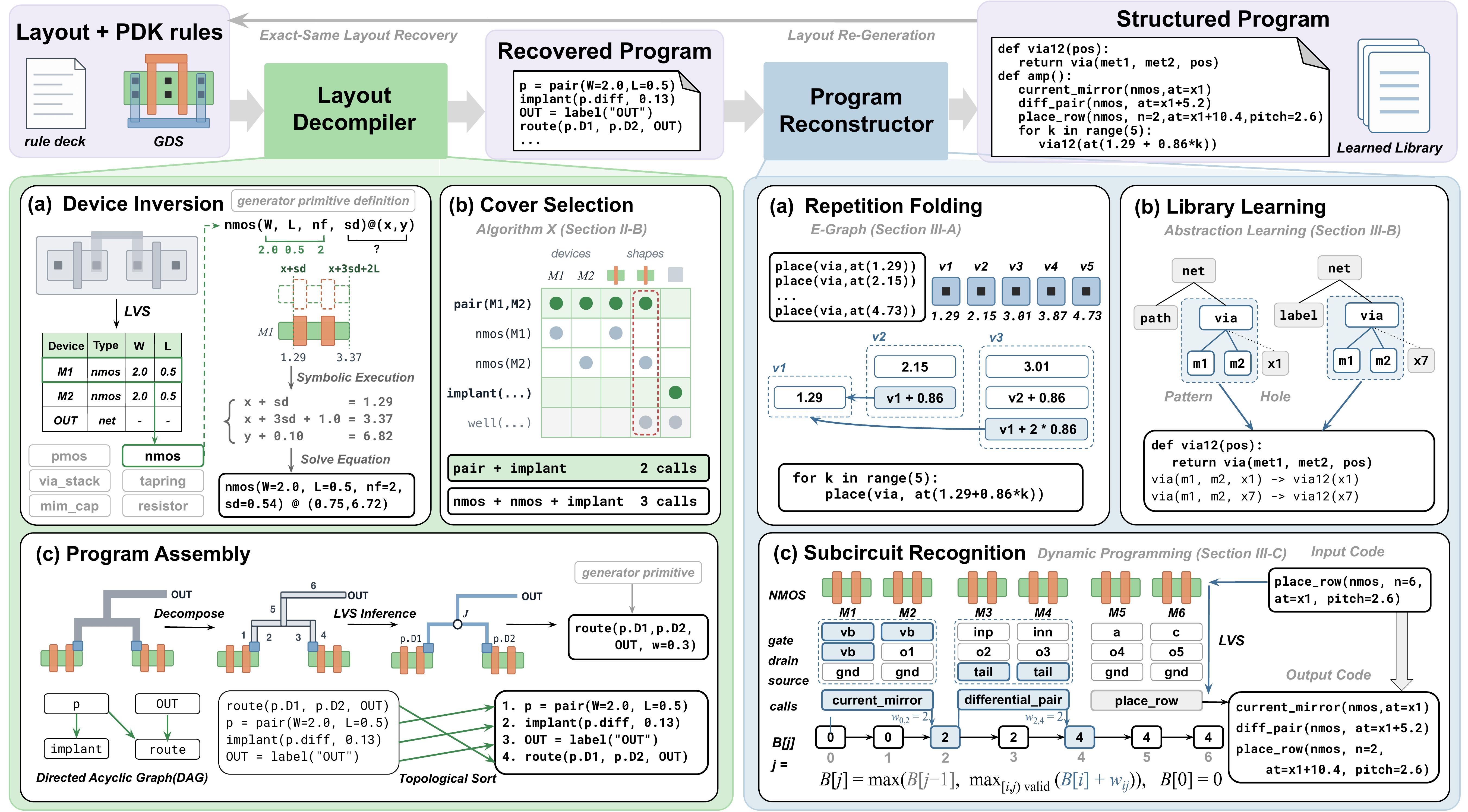}
  \vspace{-8pt}
    \caption{ReLay layout decompiler and program reconstructor. Both the flat and the structured program regenerate the original GDS byte for byte.}
  \label{fig:compiler-reconstructor}
  \vspace{-12pt}
\end{figure*}

How can a generator library cover the full variety of real designs while making
every new design more efficient to write? Our key insight is that the vocabulary of a
generator can be learned from layouts themselves. Every complete layout records
the intent of its designer, the techniques that make it work, and the trade-offs
settled while tuning it. Generalized across many layouts, these regularities
yield functions that are reusable by construction and that match real design
practice, and a library learned in this way keeps growing with each layout it
encounters. To learn from layouts, ReLay (Figure~\ref{fig:intro-compare}, bottom)
reverses the usual generator workflow, starting from a finished layout and
recovering the generator program that produces it. 

We cast this recovery as a decompilation problem and solve it with program
synthesis~\cite{SolarLezama2008Sketching}, equality
saturation~\cite{Willsey2021egg}, and library
learning~\cite{Ellis2021DreamCoder,Bowers2023Stitch}. The layout decompiler
translates any layout in a supported process into a flat program, after which
the Program Reconstructor rewrites that program into a structured generator in
which repeated units become loops, common constructions become shared functions,
and familiar circuit blocks such as current mirrors are encapsulated as modules
and called by name. Every function and argument is recovered directly from the
layout, so the program encodes the reasoning of its designer and regenerates the
original GDS file byte for byte, while earlier attempts at layout-to-generator
conversion~\cite{Jeong2024L2G,Jeong2025CNN} still leave a designer to assign
existing generators to sub-cells and write their parameters and any missing
generator by hand. Once translated into code, a layout can migrate to another process by editing one
line of configuration, arriving not as a frozen copy but as a live generator, ready
to be resized, retuned, and reused. As layouts pass through ReLay, its library
learns the constructions they share, so every layout makes the next design more
efficient to write.

 This work makes three main contributions:
\begin{itemize}
  \setlength{\itemsep}{0.15em}
  \setlength{\parskip}{0pt}
  \setlength{\parsep}{0pt}
  \item \textbf{A layout decompilation
  framework} that converts a layout in a supported process into structured, editable, human-readable Python code. With it, a layout can be regenerated in another process following the original layout style and design intent.
  \looseness=-1

  \item \textbf{A learned generator library} that grows well beyond handcrafted ones with design intent mined from reverse-engineered layouts, giving a code-to-layout generator framework that is more efficient and covers more layouts than state-of-the-art generators.\footnote{Full Code will be release upon publication}

  \item \textbf{The first layout--code benchmark,} with 1,050 human-drawn layouts in 27 circuit families, from single cells to full chips.\footnote{Benchmark Dataset will be released upon publication}
  
\end{itemize}

\vspace{-2pt}
\section{Layout Decompiler}
\label{sec:compiler}
\vspace{-2pt}

The layout decompiler (Figure~\ref{fig:compiler-reconstructor}, left) turns a layout into a \emph{flat program}, the first executable description of the layout in the generator language. Every transistor, guard
ring, implant, via, and wire becomes a call to a parametric generator, and the
values a designer sets by hand, such as channel width and finger count, become the arguments of these calls. In a flat program, each
functional shape of the layout is drawn by exactly one call. Therefore, every
polygon can be traced back to the generator call that drew it, and the line of code where that happened.
\looseness=-1

\vspace{-2pt}
\subsection{Device Inversion}
\label{sec:device-inversion}
\vspace{-2pt}

Device inversion recovers how each device was actually drawn, that is, the
generator call that produced it together with all its arguments. LVS
extraction with the foundry deck supplies part of this
information, identifying every device with its type and constituent shapes. The channel width $W$ and length $L$ come directly from LVS, and the
finger count $n_f$ follows from the gate strips that cross the active area. The
remaining arguments lie outside what extraction can see. These are the drawing dimensions,
such as the source/drain width $s_d$, and the placement $(x,y)$, which describe how and where the device was drawn.
\looseness=-1

We studied a range of layout
generators~\cite{Chang2018BAG2,Han2021LAYGO,Kunal2019ALIGN,Hammoud2024GLayoutLLM,Hammoud2024GLayoutRL}.
Although they construct devices and structures of very different forms, their
coordinate computation shares a common mathematical structure. Once the discrete
configuration of a generator is fixed, such as its finger count and the presence of
dummy fingers or a substrate tap, the generator acts as an
affine map from its parameters to its geometry: every edge it draws is a linear
combination of the unknown parameters plus a constant. Building on this property, we cast device inversion as a
program sketching problem~\cite{SolarLezama2008Sketching}. In sketching, a
programmer writes the structure of a program and leaves some of its constants
blank as \emph{holes}; a solver then fills the holes so that the completed
program meets a given specification. Device inversion fits this pattern closely: the generator supplies how every shape is constructed, leaving its unknown drawing dimensions and placement as holes for the observed shapes to determine. Filling the holes then amounts to solving a linear system, which we derive by symbolic execution~\cite{King1976SymbolicExecution}, running the unmodified source of the generator on symbols in place of the unknowns. Matching the edges it emits with the corresponding edges of the observed device yields
\looseness=-1
\begin{equation}
\label{eq:edge-system}
\mathbf{a}_k^{\top}\theta+b_k=o_k,\qquad k=1,\dots,m,
\end{equation}
where $\theta$ collects the unknown drawing dimensions and placement, and $o_k$
is the observed coordinate of the $k$-th edge. Since the equations are derived
from the generator code, the same mechanism inverts every generator in the
library, without a fitting procedure designed for each device family. Code that branches on an unknown or multiplies two unknowns is checked once the affine system has bound the dimensions it reads, and Gauss--Jordan elimination over the rationals solves the system of each discrete configuration.
\looseness=-1

The elimination has three possible outcomes. A unique solution determines every unknown dimension and the placement, and substituting them into the generator completes the call. A contradictory equation shows that the observed edges cannot all come from this generator under this configuration, and a rank-deficient system leaves some dimensions free; either way, the configuration is ruled out. Finally,
each solved call
is executed once and its full output is compared with the device, which checks
the geometry that the matched edges leave unconstrained, such as contact arrays
and enclosures; the calls that pass form the candidate set of cover selection. Beyond individual devices, material shared by several devices, such as an
implant or a well, is written as a region relation over the inverted devices,
for example an enclosure of their diffusions grown by a margin, so that it
follows them when they are resized.
\looseness=-1

\vspace{-2pt}
\subsection{Cover Selection}
\label{sec:cover-selection}
\vspace{-2pt}

Device inversion provides generator calls for every device, and the library is large enough that a device often has more than one.
For example, the two transistors of a pair can be drawn by one pair generator or by two
single-transistor generators, and the well around them can be drawn by a call of
its own or come with a generator that already includes it. Since these
alternatives share devices and shapes, the choice made for one device shapes the
options of its neighbors. Cover selection makes this choice jointly for the whole layout, assembling the per-device explanations into one consistent explanation.
\looseness=-1

We cast this choice as an exact-cover
problem~\cite{Karp1972Reducibility,Knuth2000DancingLinks}, which asks for a set
of rows of a binary matrix that covers every column exactly once. The Cover
Selection panel of Figure~\ref{fig:compiler-reconstructor} shows this matrix for
two transistors M1 and M2. Each call that device inversion returns is a
\emph{verified} candidate and forms a row. Each device and each functional
shape is a column, and a dot marks that the row's call draws that device or owns
that shape. We write $D_r$ and $O_r$ for the devices and shapes that row $r$ covers.
The matrix is the incidence matrix of a bipartite graph that links every
candidate to the devices and shapes it covers, and the selection below is a
search on this graph.
\looseness=-1

Every device column must be covered exactly once, whereas a shape column may never be covered twice but may stay uncovered, since unowned shapes are completed by material relations or routes (Section~\ref{sec:program-assembly}). As a result, two rows that share a device or a shape exclude each other, as the pair and the well candidate do over the shape in the dashed red column.
\looseness=-1

Exact cover is NP-complete in general~\cite{Karp1972Reducibility}. Our search
extends Knuth's Algorithm~X~\cite{Knuth2000DancingLinks}, a backtracking search
that builds a selection one row at a time, with the execution
feedback and the bounds described below (Algorithm~\ref{alg:cover}). For a
partial selection $S$, let $D(S)$ and $O(S)$ be the devices and shapes its rows
cover. The devices outside $D(S)$ are pending, and a candidate remains
\emph{available} as long as it shares no device or shape with the rows already
selected. At each step, the search takes the pending device with the fewest available candidates, M1 in the figure, which keeps the search tree narrow, and tries its candidates larger ones first, so that the first selections it completes use few calls.
\looseness=-1

\begin{algorithm}[t]
\caption{Exact cover selection with execution feedback}
\label{alg:cover}
\fontsize{7.3}{8.2}\selectfont
\begin{algorithmic}[1]
\REQUIRE verified candidates $R$ with devices $D_r$ and shapes $O_r$; devices $\mathcal{D}$
\ENSURE best selection $S^\star$
\STATE $X_r\gets$ the candidates that $r$ excludes, for every $r\in R$
\STATE $\beta\gets$ call bound of Eq.~\eqref{eq:call-bound};\; $S^\star\gets\emptyset$
\STATE \textsc{Visit}$(\emptyset,R,\emptyset)$;\; \textbf{return} $S^\star$
\STATE \textbf{procedure} \textsc{Visit}$(S,V,U)$ \COMMENT{selected, available, unresolved}
\STATE \quad $Q\gets\mathcal{D}\setminus(D(S)\cup U)$ \COMMENT{pending devices}
\STATE \quad \textbf{if} $|D(S)|+|Q|<|D(S^\star)|$ \textbf{or} search ended \textbf{then return}
\STATE \quad move every $d\in Q$ that no $r\in V$ covers from $Q$ to $U$
\STATE \quad \textbf{if} $Q=\emptyset$ \textbf{then} \COMMENT{a leaf}
\STATE \quad\quad $F(S)\gets$ nets that fail when the residual wires of $S$ are routed
\STATE \quad\quad \textbf{if} $\sigma(S)>\sigma(S^\star)$ \textbf{then} $S^\star\gets S$
\STATE \quad\quad \textbf{if} all devices and shapes covered, $F(S)=\emptyset$, $|S|=\beta$ \textbf{then} stop
\STATE \quad\quad \textbf{return}
\STATE \quad $d\gets$ the device in $Q$ with the fewest candidates in $V$
\STATE \quad \textbf{for all} $r\in V$ with $d\in D_r$, largest $|D_r|$ first \textbf{do}
\STATE \quad\quad \textsc{Visit}$(S\cup\{r\},\,V\setminus X_r,\,U)$
\STATE \quad \textsc{Visit}$(S,\,V,\,U\cup\{d\})$ \COMMENT{$d$ left unresolved}
\end{algorithmic}
\end{algorithm}

When no device is pending, the search reaches a leaf and checks the selected device calls by execution. Different selections expose different ports, so a cover that explains every device may still leave wires that the routing primitives cannot recover. The decompiler routes these residual wires (Section~\ref{sec:program-assembly}) and records the failed nets $F(S)$. With $O(S)$ including the routed shapes, the lexicographic key $\sigma(S)=(|D(S)|,\,|O(S)|,\,-|F(S)|,\,-|S|)$ ranks explained devices, owned shapes, failed nets, and device calls, in that order. Fewer calls are preferred only when the preceding criteria tie. Implant and well calls are recovered separately as material relations and do not count toward $|S|$.
\looseness=-1

Two bounds keep the search small. The first prunes a partial selection that cannot catch up with the best one found. The second tells the search when to stop.
Projecting the bipartite graph onto the devices links two devices whenever some
candidate draws both. Each connected component $K$ of this device graph is
a group of devices that no candidate spans across, so the group is covered by its
own calls, and since one call covers at most $\max_r|D_r\cap K|$ of its devices,
every selection needs at least
\begin{equation}
\label{eq:call-bound}
\beta=\sum_{K}
\left\lceil\frac{|K|}{\max_{r}|D_r\cap K|}\right\rceil
\end{equation}
device calls. A selection covering every device and functional shape,
with no failed nets and $\beta$ device calls, cannot be improved.
In Figure~\ref{fig:compiler-reconstructor}, \texttt{pair} reaches the
device-call bound $\beta=1$, versus two \texttt{nmos} calls;
the additional \texttt{implant} call gives the totals of two and three shown.
\looseness=-1

\vspace{-2pt}
\subsection{Program Assembly}
\label{sec:program-assembly}
\vspace{-2pt}
Wires are explained by generators as well, through the \texttt{route}
primitive. As the Program Assembly panel of Figure~\ref{fig:compiler-reconstructor} shows, the decompiler first breaks each wire into straight rectangular segments,
then chains the segments into paths between the terminals that LVS assigns to
the net, splitting a path wherever the wire branches. Each path is written in the
vocabulary of the routing generators: its turn pattern, the layer and width of
every segment, and the via stacks where it changes layers. The paths end at the Ports that the selected calls expose, so a recovered call such as \texttt{route(p.D1, p.D2, OUT, w=0.3)} is anchored to the devices it connects and follows them when they move. Since a terminal often sits inside a child cell while its wire belongs to the parent, the decompiler works bottom-up, carrying the verified metal of each child into its parent through every placement so that parent routes land on generated metal.
\looseness=-1

Finally, the calls are ordered. Up to this point, they still follow the order in which
their shapes appear in the layout, so a call can precede the calls whose ports
it reads. In the panel, the route reads both the pair and the pin, and such
reads form the directed acyclic graph shown on its left. A stable topological
sort~\cite{Kahn1962Topological} places every call after the calls it reads, which yields pair, implant, label, route.
Keeping the order stable aligns the program with the layout and makes the
programs of related layouts comparable line by line. Child cells precede their
parents in the same way, and the ordered calls form the flat program.

\vspace{-2pt}
\section{Program Reconstructor}
\label{sec:reconstructor}
\vspace{-2pt}

The flat program describes the layout literally, with each device and wire drawn by its own call at its own coordinates. However, the same literal form
leaves the design knowledge behind the layout implicit in the numbers. For
example, the finger pitch, the mirror symmetry between the two halves, and the
guard ring that encloses both together shape the layout of a differential pair.
Absolute coordinates cannot express these relations, so widening the transistors means moving every finger, contact, and ring edge by hand. The Program Reconstructor (Figure~\ref{fig:compiler-reconstructor}, right) rewrites the flat program into a \emph{structured program} that draws the same layout while being modular, readable, reusable, and easy to retarget to new sizes and processes. To do so, it first restores the cell hierarchy by turning each cell of the flat program into a function, then recovers three further kinds of design knowledge through repetition folding, library learning, and subcircuit recognition.
\looseness=-1

\vspace{-2pt}
\subsection{Repetition Folding}
\label{sec:repetition-folding}
\vspace{-2pt}

A designer who places identical units at equal pitch writes a loop, while the
flat program writes one call per unit at its own absolute position. Each position admits several descriptions, and folding the calls into a loop requires the one description that every unit shares. To avoid
committing to a description too early, we keep all of them in an
e-graph~\cite{Tate2009EqualitySaturation,Willsey2021egg}, a structure from
compiler optimization that collects expressions known to be equal into an
\emph{e-class} and stores their common parts only once. Each dashed box in the
Repetition Folding panel of Figure~\ref{fig:compiler-reconstructor} is the
e-class of one via position.
\looseness=-1

The e-graph is built over a small algebra of placements for a group of repeated
units, in which each member sits at an absolute origin, at another member's
position plus an offset, at an anchor plus a whole number of pitches along a row
or a grid, at the mirror image of another member, or, for a guard ring,
around a group of devices grown by fixed margins. Rewrite rules add to the e-class of each member every equal form that the exact integer layout coordinates confirm, until no rule finds anything new, a process known as equality saturation~\cite{Tate2009EqualitySaturation}. Extraction then
picks, for each group, the single relation that places every member with the
fewest program tokens, counting the names, numbers, and symbols of the source
code, and writes the group as a loop or, for a run of identical transistors, as a single row call.
Szalinski~\cite{Nandi2020Szalinski} folds repetitive CAD models by equality
saturation in a similar spirit, while our algebra adds the mirrored pairs and
enclosing rings that analog layout relies on. Statement blocks that still repeat are rerolled into loops by anti-unification~\cite{Plotkin1970Generalization}, which turns the values in which the blocks differ into a loop variable. Cells of one program that differ only in their values merge into one function over these values.

\vspace{-2pt}
\subsection{Library Learning}
\label{sec:library-learning}
\vspace{-2pt}

Repetition occurs not only within a program but also across programs, where the
same construction is written out in full in every design that uses it. A
designer would define such a construction once as a function and call it
wherever it is needed, and library learning~\cite{Ellis2021DreamCoder} automates
this step by searching a corpus of programs for the functions whose calls
shorten the corpus the most. 

Our search follows Stitch~\cite{Bowers2023Stitch},
which views every call as a syntax tree. A candidate library function, called an
\emph{abstraction}, is written as a \emph{pattern}, a call tree with some
arguments left blank, and each blank, called a \emph{hole}, becomes one of its
parameters. The search grows patterns from the top down. It stores the calls of all programs in a hash-consed directed acyclic graph (DAG), in which identical subtrees are kept once, so the uses of a pattern can be counted over the whole corpus. For each call shape, a callee together with the number and keywords of its arguments, the search starts from the most general pattern, the callee with every argument a hole, and refines the leftmost hole step by step. Each refinement either keeps the hole as a parameter, as \texttt{pos} in the Library Learning panel of Figure~\ref{fig:compiler-reconstructor}, or fixes it to a value that appears at that position in at least three programs, such as \texttt{met1} and \texttt{met2}. Following the minimum description length principle~\cite{Rissanen1978MDL}, a pattern $\alpha$ is worth the tokens it saves,
\begin{equation}
\label{eq:mdl-utility}
U(\alpha)=n_\alpha\,(f_\alpha-\kappa)-d_\alpha-2\,p_\alpha ,
\end{equation}
where $n_\alpha$ counts its uses, $f_\alpha$ the tokens it fixes in each use,
$\kappa$ the tokens of a call, $d_\alpha$ the tokens of its definition, and
$p_\alpha$ the programs that import it, each import costing two tokens. Since a refinement can fix more tokens but never gains uses, the utility that $\alpha$ would reach if every open hole were fixed while all its current uses remained, $\hat{U}(\alpha)$, bounds every refinement of $\alpha$, and Algorithm~\ref{alg:library}, a best-first search on $\hat{U}$, stops as soon as no open pattern can beat the best one found.

Two constraints adapt the search to layout programs. Every hole must be free of
calls, since an abstraction evaluates its arguments before its body, and a
generator call inside an argument would run ahead of the calls in the body and change the drawing order. For readability, an abstraction
takes at most eight parameters and must be used by at least three programs from
at least three independent design groups, so that it records practice shared
across designers. The learned functions form the top layer of the library, above the built-in primitives and the generator extensions that the decompiler relies on.
\looseness=-1

\begin{algorithm}[t]
\caption{Top-down library learning}
\label{alg:library}
\fontsize{7.3}{8.2}\selectfont
\begin{algorithmic}[1]
\REQUIRE corpus $\mathcal{P}$ of programs
\ENSURE library $\Lambda$ and the rewritten corpus
\STATE $\mathcal{H}\gets$ hash-consed DAG of the calls in $\mathcal{P}$;\; $\Lambda\gets\emptyset$
\REPEAT[one abstraction per round]
  \STATE $\alpha^\star\gets\bot$ with $U(\bot)=0$
  \STATE $\mathcal{Q}\gets$ the most general pattern of every call shape in $\mathcal{H}$
  \WHILE[prune by $\hat{U}$]{some $\alpha\in\mathcal{Q}$ has $\hat{U}(\alpha)>U(\alpha^\star)$}
    \STATE pop the pattern $\alpha$ with the largest $\hat{U}(\alpha)$ \COMMENT{best-first}
    \IF{$\alpha$ meets both constraints \AND $U(\alpha)>U(\alpha^\star)$}
      \STATE $\alpha^\star\gets\alpha$
    \ENDIF
    \STATE push the refinements of the leftmost hole of $\alpha$ \COMMENT{keep or fix}
  \ENDWHILE
  \IF[later rounds build on it]{$\alpha^\star\neq\bot$}
    \STATE $\Lambda\gets\Lambda\cup\{\alpha^\star\}$;\; rewrite every use of $\alpha^\star$ in $\mathcal{H}$
  \ENDIF
\UNTIL{$\alpha^\star=\bot$}
\RETURN $\Lambda$
\end{algorithmic}
\end{algorithm}

\vspace{-2pt}
\subsection{Subcircuit Recognition}
\label{sec:subcircuit-recognition}
\vspace{-2pt}

Repetition folding groups devices by their
placement, so a run of identical transistors at one pitch becomes a single row call.
Since placement alone decides this grouping, one row call can hold devices of
different circuit functions, such as a current mirror beside a differential pair, and
the program leaves the circuit structure of the row unstated. Subcircuit recognition
recovers this structure by splitting each row call into calls of the subcircuits it
contains. Recognition reads the netlist that LVS extracts. As in subcircuit
identification~\cite{Ohlrich1993SubGemini,Kunal2020GANA}, each subcircuit is defined by
a rule over device types, sizes, terminal nets, and positions, from two-transistor
pairs through cascode mirrors and matched arrays to guard rings that enclose whole
groups of devices.
\looseness=-1

Choosing which subcircuits to write is harder than finding them. Candidate
subcircuits overlap, as when the two transistors of a mirror also belong to a longer
array of identical transistors, while each device plays a single role, so the
reconstructor must select disjoint candidates that together cover the most devices.
Over arbitrary groups of devices this selection is weighted set packing, an NP-hard
problem~\cite{Karp1972Reducibility}. However, the row call supplies the structure
that makes it tractable. The row orders its members, and this order turns every
usable candidate into an interval of consecutive members. Selecting disjoint
intervals of largest total weight on a line has optimal substructure, because the
best cut of the first $j$ members depends on nothing to its right. With $B[j]$ the number of devices that the best cut of the first $j$ members places in subcircuits, and $w_{ij}$ the number of devices in a valid run $[i,j)$, whose members $i$ to $j-1$ draw one subcircuit, this gives the dynamic program~\cite{Bellman1957DP}
\looseness=-1
\begin{equation}
\label{eq:segment-dp}
B[j]=\max\Big(B[j-1],\ \max_{[i,j)\ \text{valid}}\big(B[i]+w_{ij}\big)\Big),
\qquad B[0]=0.
\end{equation}
The recurrence finds the optimal cut exactly in $O(n^2)$ time for a row of $n$
members, and the scan from each start $i$ stops at the first member that draws a
device of no subcircuit, since no longer run through that member can become valid.
Tracing back the maximizing terms recovers the cut, as Figure~\ref{fig:compiler-reconstructor} shows on a row of six transistors.
\looseness=-1

Each run of the optimal cut becomes one call to the generator of its subcircuit.
Subcircuits spread over child cells, such as a latch built from two cross-coupled
inverter cells, are recognized over the cell hierarchy in the same way. Larger blocks such as an OTA are recognized from the subcircuits they contain. A block
belongs to the deepest cell that holds all its devices, where the statements that draw
it become one function marked with the name of the block. As a result, the structured
program takes the form of a handcrafted generator, with one named call for every
subcircuit and block.
\looseness=-1

\section{Library Evolution and Benchmark Construction}
\vspace{-2pt}

We build the benchmark from layouts in public repositories of three open PDKs,
SkyWater SKY130, GlobalFoundries GF180MCU, and IHP SG13G2, and cover as much of
the open-silicon ecosystem as we can: Tiny Tapeout shuttles and multi-project
wafer runs such as the IHP open MPW and the Caravel shuttles, design contests
such as the IEEE SSCS Chipathon 2023--2026 and UNIC-CASS, research groups,
open IP libraries, layout
datasets~\cite{Chiari2026THEIA,Li2026ICLayoutBench}, and more.

\begin{figure}[!t]
  \centering
  \includegraphics[width=\columnwidth]{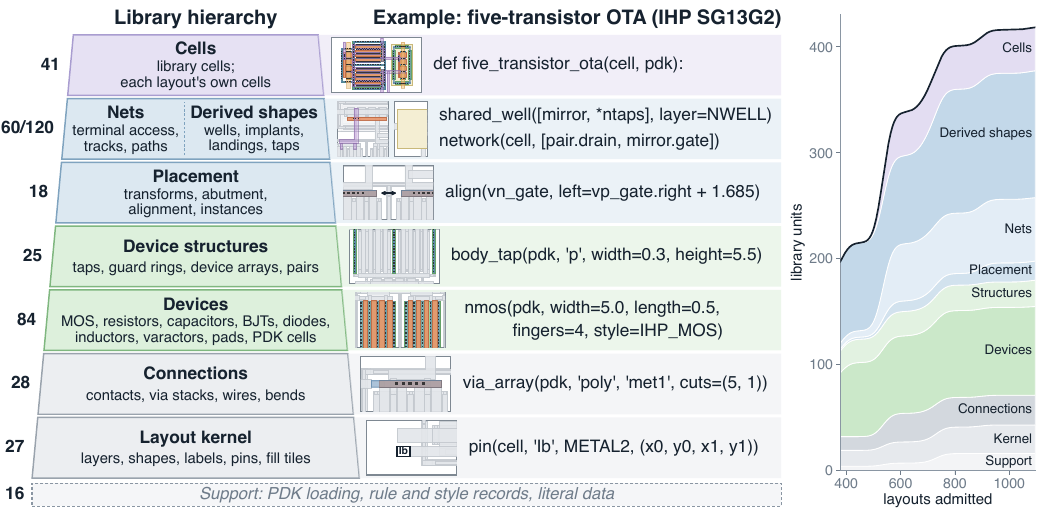}
  \vspace{-18pt}
  \caption{Library hierarchy and its evolution with the benchmark.}
  \label{fig:library-evolution}
  \vspace{-12pt}
\end{figure}

Each entry pairs a layout with its extracted netlist, a description and origin, its flat program, and its structured program. The structured programs share a 419-unit library of built-in primitives,
generator extensions, and learned functions. Extensions cover previously
unsupported constructions; library learning
(Section~\ref{sec:library-learning}) adds patterns shared across layouts. Its units are tiered by what they construct, from the
layout kernel up to large top-level cells, as shown in Figure~\ref{fig:library-evolution}. The right panel traces how the library grows as layouts accumulate. Recovery is held to the strictest criterion, where a program must regenerate the
original GDS byte for byte and its output must extract to the same netlist as the
original. Custom layouts often hold a few electrically sound shapes that follow no
template, such as a wire of an unusual width or a pad trimmed by hand, and we
tolerate such shapes written as literal coordinates in at most 0.1\% of the
records.
\looseness=-1

The benchmark holds 1,050 layouts in 27 circuit families, from amplifiers,
comparators, and oscillators to data converters, PLLs, bandgap references, LNAs,
regulators, and I/O cells; from small cells to test chips with pad rings.
ReLay supports layouts of up to 62\,MB of GDS, 1.68
million records, and 36,905 devices. Compared with the benchmarks in
Table~\ref{tab:benchmarks}, our benchmark is built entirely from human-drawn layouts and spans the most analog circuit families.

Retargeting multiplies the value of the benchmark. Every structured program names its PDK in a single line, and changing that line regenerates the layout in another PDK. ReLay redraws each device under the rules of the new PDK at the size the program gives, then lets a linear program move the shapes only as far as the new rules require while keeping which shapes lie to the left, to the right, above, or below one another, so the placement, the routes, and the arrangement of matched devices carry over unchanged. Thus, each layout can yield up to two DRC- and LVS-clean counterparts in the other PDKs, regenerated by the same program, which turns the benchmark into matched sets of layouts across processes for studying process-portable design.
\looseness=-1

\section{Evaluation}
\label{sec:eval}
\vspace{-2pt}

\vspace{-2pt}
\subsection{Recovery of Held-Out Layouts}
\label{sec:held-out}
\vspace{-2pt}

\begin{figure*}[t]
  \centering
  \includegraphics[width=\textwidth]{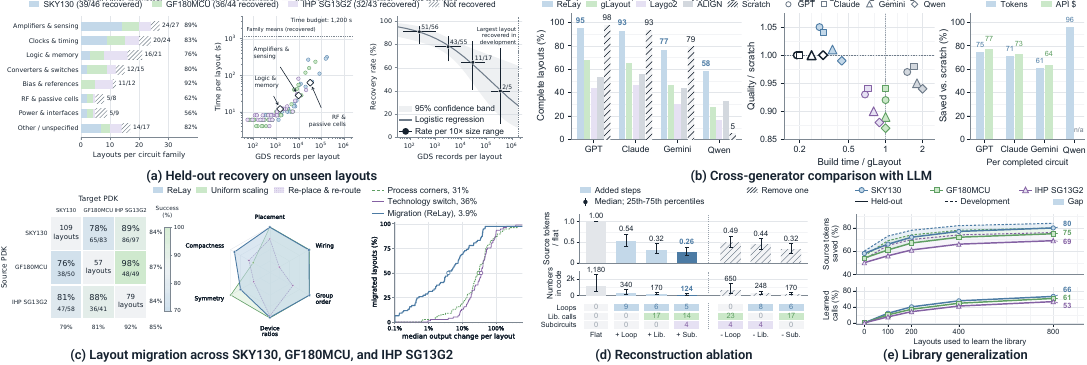}
  \vspace{-15pt}
    \caption{Evaluating ReLay on multiple tasks. In (b), cost counts all tokens and API charges per completed task, including failed attempts, and build time covers regenerating a completed layout; in (d) and (e), program size counts source tokens, including library definitions. In (e), learning uses only development layouts.}
  \label{fig:evaluate-compare}
  \vspace{-9pt}
\end{figure*}

We held out 133 benchmark layouts during development and ran the frozen decompiler and reconstructor on each of them once. The decompiler recovers 107 of them exactly, in a median of 8.1\,s, and the reconstructor restructures every recovered program (Fig.~\ref{fig:evaluate-compare}(a)). Apart from three decompiler defects, every failure has one cause: the layout holds a construction that the library cannot yet express, such as a 5\,V transistor with extra implants, the tap rows of a flattened standard cell, or a via array repeated in 64 rows. Because a program is accepted only when it regenerates the layout exactly, such a gap stops recovery rather than yielding a wrong program. The cause also explains the trends in the figure: a larger layout is more likely to hold one unseen construction, so recovery falls with size, and bias and reference circuits, built from standard devices, recover best, whereas power and interface circuits, which rely on high-voltage and ESD devices, recover worst. Thus, each failure marks a construction for the library to learn next rather than a limit of the search.
\looseness=-1

\begin{table}[t]
  \centering
  \caption{Comparison with published analog circuit benchmarks.}
  \label{tab:benchmarks}
  \vspace{-8pt}
  \fontsize{7.4}{8.7}\selectfont
  \setlength{\tabcolsep}{1.5pt}
  \begin{tabular*}{\columnwidth}{@{\extracolsep{\fill}}lrcccc@{}}
    \toprule
    Benchmark & Entries & Families & PDK & Layout & Verification \\
    \midrule
    FALCON~\cite{Mehradfar2025FALCON} & 20 & 5 & 45\,nm & none & sim. \\
    AnalogGym~\cite{Li2024AnalogGym} & 30 & 5 & SKY130 & none & sim. \\
    LOADBench~\cite{Leibl2026LOADBench} & 100+ & 1 & SKY130 & none & sim. \\
    Analog-DB~\cite{NooriZadeh2026AnalogDB} & 68 & 16 & 3 open$^{*}$ & none & sim. \\
    SGAnalog~\cite{Li2026SGAnalog} & 273 & 7 & SKY130 & none & graph, sim. \\
    OSIRIS~\cite{Chiari2026OSIRIS} & 87,100 & 2 & SKY130 & generated & DRC, LVS, sim. \\
    THEIA~\cite{Chiari2026THEIA} & 30,031 & 4 & SKY130 & generated & DRC, LVS \\
    \midrule
    \textbf{Ours} & \textbf{1,050} & \textbf{27} & \textbf{3 open}$^{*}$ & \textbf{human-drawn} & \textbf{DRC, LVS, sim.} \\
    \bottomrule
  \end{tabular*}\par\vspace{1pt}
  \makebox[\columnwidth][l]{sim.: simulation; graph: netlist graph matching. $^{*}$SKY130, GF180, SG13G2.}
  \vspace{-22pt}
\end{table}

\begin{table}[t]
  \centering
  \caption{Design properties compared before and after migration.}
  \label{tab:intent}
  \vspace{-8pt}
  \fontsize{7.4}{8.7}\selectfont
  \setlength{\tabcolsep}{3pt}
  \begin{tabularx}{\columnwidth}{@{}lX@{}}
    \toprule
    Property & Share of the source that the migrated layout keeps \\
    \midrule
    Placement & finger pairs that keep their left/right and above/below order \\
    Wiring & wire length kept on the same net, metal layer, and direction \\
    Group order & matched groups (common-centroid, interdigitated, or side-by-side) whose fingers keep their order \\
    Device ratios & matched and mirrored pairs whose $W/L$ ratio stays within 1\% \\
    Symmetry & device groups mirrored about an axis, such as the two halves of a differential pair, that stay symmetric within 5\,nm \\
    Compactness & area, as the source area over the migrated area (at most 1) \\
    \bottomrule
  \end{tabularx}
  \vspace{-15pt}
\end{table}

\vspace{-2pt}
\subsection{The Learned Library as an Agent Toolbox}
\vspace{-2pt}
\label{sec:eval-agents}

A learned library proves its value in the new layouts written with it, and LLM agents are natural writers of such code: a single call to a learned function replaces many geometric decisions, so an agent can devote its reasoning to selecting and composing constructions~\cite{Chu2026CodeBehindLayout}. We test this with four agents, GPT-6.1 Sol, Claude Opus 5.5, Gemini~4 Hydrogen, and Qwen3.6-35B-A3B, an open-weight model that runs on a single H100 GPU and incurs no API cost. Each agent solves 43 layout design tasks from 21 public repositories, released with our code: open-source circuits collected mainly through SGAnalog~\cite{Li2026SGAnalog} and example designs of ALIGN~\cite{Kunal2019ALIGN} and Laygo2~\cite{Han2021LAYGO}. A task specifies the device types, sizes, terminal connections, and pins of a published circuit, from a one-transistor bias generator to a 300-unit amplifier, and asks for a new layout without providing any reference layout.
\looseness=-1

Within each agent, only the toolbox varies: ReLay's learned library, gLayout~\cite{Hammoud2024GLayoutLLM,Hammoud2024GLayoutRL}, Laygo2, ALIGN with its netlist-driven placement and routing, or Scratch, in which the model writes the layout geometry directly. To keep the comparison fair, every toolbox comes with its documentation and a starter example of equal depth, which shows how to read a task and emit the output but performs no placement or routing, and every trial runs in an isolated Harbor container~\cite{Harbor2025} under the same interaction budget and checker feedback, without network access or access to other trials. For each agent, electrical quality is evaluated on circuits completed by all methods, with device sizes and operating conditions fixed and wire resistance and capacitance extracted from each layout. Quality is the geometric mean of performance ratios to Scratch, oriented so that higher is better, with equal weight per design source.
\looseness=-1

Figure~\ref{fig:evaluate-compare}(b) shows three findings. First, ReLay is the most effective generator library: for every agent it yields the highest rate of DRC- and LVS-clean layouts, 95\%, 93\%, 77\%, and 58\%, exceeding the next-best library by 25 to 31 percentage points, and it builds layouts in less than half the time of gLayout. Second, it substantially reduces the cost of frontier agents without degrading quality. Writing geometry directly, these agents complete a comparable fraction of the tasks (98\%, 93\%, and 79\%), but ReLay reduces their tokens per completed task by 61--75\% and their API cost by 64--77\%, while matching the electrical quality of their scratch layouts (0.99--1.05), which the other libraries do not (0.87--0.98). Third, it makes layout design attainable for an open-weight model: Qwen3.6-35B-A3B completes only 5\% of the tasks by writing geometry directly but 58\% with ReLay. 

\vspace{-2pt}
\subsection{Cross-PDK Migration}
\label{sec:eval-transfer}
\vspace{-2pt}

We evaluate retargeting on 245 recovered layouts that were split off at random before the retargeting step was tuned and were migrated only once, to the other two PDKs by their one-line edit; layouts with devices that have no counterpart in the target PDK are left out. A migration is verified when it is DRC-clean, matches the source in LVS including bulk connections, and keeps every device at its drawn size. Most migrations are verified (Fig.~\ref{fig:evaluate-compare}(c), left). The failures arise where the PDKs differ in more than dimensions: resistors and capacitors, whose bodies, contacts, and markers each PDK draws in its own way, cause most LVS mismatches, and the remaining layouts violate rules that the source process never had, such as the well rules of SKY130 or the dual-voltage rules of GF180. They reflect the gap between processes rather than the migration itself.
\looseness=-1

Design intent cannot be read off a single number, so we pair the transistor fingers of the source and the migrated layout through their extracted netlists and measure how much of each design choice survives (Table~\ref{tab:intent}). We compare ReLay with uniform scaling, which enlarges the layout by the smallest integer factor that clears the target's rules, and with re-placing and re-routing the netlist, as netlist-driven tools do, on the layouts between SKY130 and GF180 that all three can migrate (Fig.~\ref{fig:evaluate-compare}(c), middle). Uniform scaling keeps every relation only by growing the layout two to three times per side, and re-placement discards the source arrangement altogether. ReLay keeps nearly every relation at close to the source area, because its linear program preserves the relative order of all shapes (Section~IV).
\looseness=-1

Moving a circuit to another process changes its performance even when nothing else changes, since supplies and device models differ. To isolate what the migration adds, we simulate each circuit at the typical corner as the source schematic and layout in the source PDK and as the same schematic and the migrated layout in the target PDK: the schematic ratio measures the technology switch, and the layout ratio divided by it measures the migration. On 77 migrated layouts with testbenches, the migration adds a median 3.9\% to the outputs, nearly an order of magnitude below both the technology switch and the target's process-corner spread (Fig.~\ref{fig:evaluate-compare}(c), right), because every device keeps its size and only the wiring changes. A designer can therefore retarget first and re-size only where the new process demands it.
\looseness=-1

\subsection{Reconstructor Ablation}
Each reconstruction step removes a different kind of redundancy (Fig.~\ref{fig:evaluate-compare}(d)). Adding repetition folding, learned functions, and named subcircuits in turn shrinks the median program to 26\% of its flat size and cuts the numbers written in code from 1,180 to 124. Removing one step at a time shows where the savings come from. Without folding, programs nearly double and keep five times as many numbers, since loops absorb the coordinates of repeated units; without learned functions, they grow by 70\%, since shared constructions are spelled out again; without subcircuits, they grow least, as this step names device groups rather than removing code. The library generalizes (Fig.~\ref{fig:evaluate-compare}(e)): on held-out layouts, it saves 69--80\% of tokens, close to development savings, and supplies 53--66\% of layout-building calls.
\looseness=-1

\vspace{-2pt}
\section{Conclusion, Discussion, and Limitations}
\vspace{-2pt}

ReLay translates analog layouts and generator code in both directions and evolves its library from the layouts it translates. From 1,050 human-drawn layouts in three open PDKs, it delivers a layout--code benchmark and a generator library built on the collective practice of open-source designers, bringing reusable and process-portable analog design within reach. However, ReLay learns only from the layouts that the open-source community publishes, and such layouts remain scarce, since the first open PDKs appeared only a few years ago; therefore, the corpus draws on every public analog layout we could locate. We reviewed each by hand, kept only distinct ones, and excluded test structures and visibly broken or malformed circuits; layouts whose authors provide testbenches were also checked in simulation. Even so, quality rests with each designer, and the electrical performance of some layouts still falls short of a careful design. Such layouts keep their value, because ReLay turns each into an editable program whose sizes, placements, and routes a designer can tune, so the layout serves both as a benchmark entry and as a starting point for development. We are dedicated to maintaining and extending ReLay and its benchmark as the open-source analog ecosystem grows, so that every newly published layout, even an imperfect one, can enrich the library on which future designs are built.
\looseness=-1

\bibliographystyle{IEEEtran}
\balance
\bibliography{reference}
\end{document}